\documentclass[aps,pra,showpacs,11pt]{revtex4}
\usepackage[dvips]{graphics}
\begin{document}
\title{Unambiguous Discrimination Between Linearly Dependent States With Multiple Copies}
\author{Anthony Chefles}
\affiliation{Department of Physical Sciences, University of
Hertfordshire,
       Hatfield AL10 9AB, Herts, UK \\}

\begin{abstract}
A set of quantum states can be unambiguously discriminated if and
only if they are linearly independent.  However, for a linearly
dependent set, if $C$ copies of the state are available, then the
resulting $C$ particle states may form a linearly independent set
and be amenable to unambiguous discrimination. We obtain one
necessary and one sufficient condition for the possibility of
unambiguous discrimination between $N$ states given that $C$
copies are available and that the single copies span a $D$
dimensional space. These conditions are found to be identical for
qubits.  We then examine in detail the linearly dependent trine
set. The set of $C>1$ copies of each state is a set of linearly
independent lifted trine states. The maximum unambiguous
discrimination probability is evaluated for all $C>1$ with equal a
priori probabilities.

\end{abstract}
\maketitle

\section{Introduction}
\renewcommand{\theequation}{1.\arabic{equation}}
\setcounter{equation}{0} Much of the fascination with the
information-theoretic properties of quantum systems derives from
collective phenomena and processes.  On one hand, the information
contained in entangled quantum systems is of a collective nature
and is central to many intriguing applications of quantum
information, such as teleportation and quantum computing. On the
other, there are collective operations, such as collective
measurements on several quantum systems. Broadly speaking,
collective measurements on a set of systems can yield more, or
better information than one can obtain by carrying out separate
measurements on the individual subsystems, even if these are not
entangled.  For example, the use of collective measurements is
essential for attaining the true classical capacity of a quantum
channel\cite{CCapacity}, since capacities attained with receivers
performing collective measurements on increasingly large strings
of signal states are superadditive.

A further illustration of the superiority of collective over
individual measurements is the `nonlocality without entanglement'
discovered by Bennett et al\cite{NWE}.  This refers to the fact
that one can construct a set of orthogonal product states which
can only be perfectly distinguished by a collective measurement.

In this paper, we provide a further demonstration of the increased
knowledge that can be obtained using collective rather than
individual measurements, relating to unambiguous state
discrimination\cite{Review}.  Such measurements can reveal, with
zero probability of error, the state of a quantum system, even if
the possible states are nonorthogonal.  Perfect discrimination
between nonorthogonal states is impossible and the price we pay is
the non-zero probability of inconclusive results.

It has been established that unambiguous discrimination is
possible only for linearly independent states\cite{Linear}.
However, suppose that the possible states form a linearly
dependent set, but we have $C>1$ copies of the actual state at our
disposal. Unambiguous discrimination is impossible using separate
measurements on the individual copies. If, however, the possible
$C$ particle states form a linearly independent set, then
unambiguous discrimination will be possible by carrying out a
collective measurement on all $C$ copies.

In section II, we derive one necessary and one sufficient
condition for $N$ states to be amenable to unambiguous
discrimination, given that $C$ copies of the state are available
and that the possible single copy states span a finite, $D$
dimensional space.  For qubits ($D=2$), these conditions are
identical. In section III, we work out in detail a specific
example, that of multiple copies of the so-called trine states.
The trine set is linearly dependent, although the set comprised of
multiple copies of these states is linearly independent for
$C{\geq}2$. Indeed, these states are the lifted trine states
recently discussed in a different but related context by
Shor\cite{Shor}.  We obtain the maximum discrimination probability
for these multi-trine states with equal a priori probabilities and
find that it has some curious, unexpected features.

\section{Bounds on the maximum number of distinguishable states}
\renewcommand{\theequation}{2.\arabic{equation}}
\setcounter{equation}{0} Consider the following scenario: a
quantum system is prepared in one of the $N$ pure states
$|{\psi}_{j}{\rangle}$, where $j=1,{\ldots},N$.   These states are
nonorthogonal and we would like to determine which state has been
prepared.  If we are unwilling to tolerate errors, then we should
adopt an unambiguous discrimination strategy.  Such a measurement
will have $N+1$ outcomes; $N$ of these correspond to the possible
states and a further outcome gives inconclusive results.  It has
been established that the zero errors constraint leads to a
nonzero probability of inconclusive results for nonorthogonal
states\cite{Linear}.

Suppose that the $|{\psi}_{j}{\rangle}$ span a $D$ dimensional
Hilbert space ${\cal H}$.  Clearly, $D{\leq}N$.  If $D=N$, then
the states are linearly independent.  If, on the other hand,
$D<N$, then they are linearly dependent.  Whether or not the set
is linearly independent is crucial, since linear independence is
the necessary and sufficient condition for a set of pure states to
be amenable to unambiguous discrimination\cite{Linear}.

If, however, instead of having just one copy of the state, we have
$C>1$ copies, that is, one of the states
$|{\psi}_{j}{\rangle}^{{\otimes}C}$, then there is the possibility
that, even if $\{|{\psi}_{j}{\rangle}\}$ is a linearly dependent
set, $\{|{\psi}_{j}{\rangle}^{{\otimes}C}\}$ may be linearly
independent, making unambiguous discrimination possible. It is of
interest to determine the conditions under which this is so. Here,
we will obtain two general results relating to the number of
states that can be unambiguously discriminated, given that the
single copies span a $D$ dimensional space and that $C$ copies of
the state are available.  Firstly, we will show that the number of
states which can be unambiguously discriminated satisfies the
inequality
\begin{equation}
N{\leq}{C+D-1 \choose C}.
\end{equation}
To see why, let us denote by ${\cal H}_{SYM}$ the symmetric
subspace of ${\cal H}^{{\otimes}C}$.  The states
$|{\psi}_{j}{\rangle}^{{\otimes}C}$ are invariant under any
permutation of the states of the single copies and thus lie in
${\cal H}_{SYM}$. Denoting by $D_{SYM}$ the dimension of ${\cal
H}_{SYM}$, it can be shown that\cite{Bhatia}
\begin{equation}
D_{SYM}={C+D-1 \choose C}.
\end{equation}
The $|{\psi}_{j}{\rangle}^{{\otimes}C}$ will be linearly dependent
if $N$ is greater than the dimension of ${\cal H}_{SYM}$.  This,
together with Eq. (2.2), leads to inequality (2.1), which is a
necessary condition for unambiguous discrimination between $N$
states spanning a $D$ dimensional space given $C$ copies of the
state.

This bound holds for all pure states.  It is tight, in the sense
that for all $C,D$, there exists a set of $N$ pure states
$\{|{\psi}_{j}{\rangle}\}$ such that the equality in (2.1) is
satisfied and the set $\{|{\psi}_{j}{\rangle}^{{\otimes}C}\}$ is
linearly independent.  To prove this, we make use of the fact that
${\cal H}_{SYM}$ is the subspace of ${\cal H}^{{\otimes}C}$
spanned by the states $|{\psi}{\rangle}^{{\otimes}C}$, for all
$|{\psi}{\rangle}{\in}{\cal H}$.  The set of states
$\{|{\psi}{\rangle}^{{\otimes}C}\}$ is linearly dependent.
However, every linearly dependent set spanning a vector space
${\cal V}$ has a linearly independent subset which is a basis for
${\cal V}$\cite{Broida}.  Let
$\{|{\psi}_{j}{\rangle}^{{\otimes}C}\}$ be such a subset of
$\{|{\psi}{\rangle}^{{\otimes}C}\}$ for ${\cal V}={\cal H}_{SYM}$.
These states are linearly independent and satisfy the equality in
(2.1) since $N=D_{SYM}$.

We now show that any $N$ distinct pure states can be unambiguously
discriminated if
\begin{equation}
N{\leq}C+D-1.
\end{equation}
Here, the elements of the set $\{|{\psi}_{j}{\rangle}\}$ are
considered distinct iff
$|{\langle}{\psi}_{j'}|{\psi}_{j}{\rangle}|<1\;{\forall}\;j{\neq}j'$.
It will suffice to show that if $N=C+D-1$, then the states
$|{\psi}_{j}{\rangle}^{{\otimes}C}$ are linearly independent.  To
see why, we simply note that if this can be shown, then our more
general claim will be true as a consequence of the fact that any
subset of a linearly independent set is also linearly independent.

To prove that inequality (2.3) is a sufficient condition for
unambiguous discrimination, we assume that $N=C+D-1$ and again
make use of the fact that any linearly dependent set has a
linearly independent spanning subset. The set
$\{|{\psi}_{j}{\rangle}\}$ then has a subset of $D$ linearly
independent states, which we shall denote by $S_{LI}^{1}$. Without
loss of generality, we can relabel all states according to the
index $j$ in such a way that $|{\psi}_{j}{\rangle}{\in}S_{LI}^{1}$
for $j=1,{\ldots},D$.

Let us now consider the sets
$S_{LI}^{r}=\{|{\psi}_{j}{\rangle}^{{\otimes}r}|j=1,{\ldots},D+r-1\}$,
for $r=1,{\ldots},C$.  Notice that $S_{LI}^{1}$ accords with our
previous definition and that
$S_{LI}^{C}=\{|{\psi}_{j}{\rangle}^{{\otimes}C}\}$. We will use
induction to prove that $\{|{\psi}_{j}{\rangle}^{{\otimes}C}\}$ is
linearly independent. The set $S_{LI}^{1}$ is linearly independent
by definition.  We will show that if $S_{LI}^{r-1}$ is linearly
independent, then so is $S_{LI}^{r}$. To do this, we shall require
the following: \\ *

\noindent \bf {Lemma}. {\it  Let
$\{|{\chi}_{k}{\rangle}\}{\in}{\cal H}$ and
$\{|{\phi}_{k}{\rangle}\}{\in}{\cal H}'$ be sets of distinct,
normalised state vectors which have equal cardinality. Consider
any normalised states $|{\chi}{\rangle}{\in}{\cal H}$ and
$|{\phi}{\rangle}{\in}{\cal H}'$ such that $|{\chi}{\rangle}$ is
distinct from all elements of $\{ |{\chi}_{k}{\rangle}\}$. If the
set $\{|{\phi}_{k}{\rangle}\}$ is linearly independent, then so is
the set
$\{|{\phi}_{k}{\rangle}{\otimes}|{\chi}_{k}{\rangle}\}\cup(|{\phi}{\rangle}{\otimes}|{\chi}{\rangle})$.}
\\ * \rm

A proof of this is given in the appendix.  The linear independence
of $S_{LI}^{r-1}$ can be seen to imply that of $S_{LI}^{r}$ if we
make the identifications:
\begin{eqnarray}
\{|{\phi}_{k}{\rangle}\}&=&S_{LI}^{r-1}, \\ *
\{|{\chi}_{k}{\rangle}\}&=&\{|{\psi}_{j}{\rangle}|j=1,{\ldots},D+r-2\},
\\ *
|{\phi}{\rangle}&=&|{\psi}_{D+r-1}{\rangle}^{{\otimes}r-1}, \\ *
|{\chi}{\rangle}&=&|{\psi}_{D+r-1}{\rangle},
\end{eqnarray}
for $r=2,{\ldots},C$.    Thus, the set
$S_{LI}^{C}=\{|{\psi}_{j}{\rangle}^{{\otimes}C}\}$ is linearly
independent and this completes the proof. We have shown that (2.3)
is a sufficient condition for unambiguous discrimination between
$N$ states spanning a $D$ dimensional space given $C$ copies of
the state.

Like the necessary condition in (2.1), this bound is the tightest
we can obtain using $N,C$ and $D$ alone, in the sense that for all
values of these parameters which do not satisfy (2.3), there
exists a set of states $\{|{\psi}_{j}{\rangle}^{{\otimes}C}\}$
which is linearly dependent. To prove this, suppose that for
$j=1,{\ldots},D$, the $|{\psi}_{j}{\rangle}$ are linearly
independent and that for $j=D+1,{\ldots},N$,
$|{\psi}_{j}{\rangle}=a_{j}|{\psi}_{D-1}{\rangle}+b_{j}|{\psi}_{D}{\rangle}$,
for some complex coefficients $a_{j},b_{j}$.  If (2.3) is not
satisfied, then $N{\geq}C+D$ and the subspace spanned by
$|{\psi}_{D-1}{\rangle}$ and $|{\psi}_{D}{\rangle}$ contains at
least $C+2$ states in the set $\{|{\psi}_{j}{\rangle}\}$.  We will
now see that the set of states
$\{|{\psi}_{j}{\rangle}^{{\otimes}C}|j=D-1,D,{\ldots},N\}$ is
linearly dependent.  For $j=D-1,{\ldots},N$, the
$|{\psi}_{j}{\rangle}$ all lie in the same two dimensional
subspace, so that the corresponding $C$-fold copies
$|{\psi}_{j}{\rangle}^{{\otimes}C}$ lie in the symmetric subspace
of $C$ qubits, which, from Eq. (2.2), is $C+1$ dimensional.  It
 follows that if there are at least $C+2$ of these
states, they must be linearly dependent.  This implies that the
entire set of $C$-fold copies
$\{|{\psi}_{j}{\rangle}^{{\otimes}C}|j=1,{\ldots},N\}$ is linearly
dependent.

The necessary and sufficient conditions, (2.1) and (2.3), for the
linear independence of $C$ copies of $N$ states, with single copy
Hilbert space dimension $D$ are thus the most complete statements
that can be made about the possibility of unambiguous
discrimination given only these three parameters. These two bounds
are also, in general, different from each other, which implies
that for a particular set of states, additional, more detailed
information about the set may be useful.

However, this is not the case for $D=2$. For the case of qubits,
these bounds are identical and equal to $C+1$. The necessary and
sufficient condition for the possibility of unambiguous
discrimination between $N$ pure, distinct states of a qubit, given
$C$ copies of the state, is then
\begin{equation}
N{\leq}C+1.
\end{equation}
The generality of this result is quite remarkable, since it is
completely independent of the actual states involved.  These will,
however, have a strong bearing on the maximum probability of
success.
\section{Discrimination between multi-trine states}
\renewcommand{\theequation}{3.\arabic{equation}}
\subsection{Trine and lifted trine states}
\setcounter{equation}{0} Having discussed in the preceding section
the conditions under which unambiguous discrimination between
multiple copies of linearly dependent states is possible, let us
examine in detail one particular example, that of the so-called
`trine' set.  Consider a qubit whose two dimensional Hilbert space
is denoted by ${\cal H}_{2}$.  Let $\{|x{\rangle},|y{\rangle}\}$
be an orthonormal basis for ${\cal H}_{2}$.  Then the following
states form the trine set:
\begin{eqnarray}
|t_{1}{\rangle}&=&|y{\rangle}, \\ *
|t_{2}{\rangle}&=&\frac{1}{2}(-|y{\rangle}+\sqrt{3}|x{\rangle}),
\\
* |t_{3}{\rangle}&=&\frac{-1}{2}(|y{\rangle}+\sqrt{3}|x{\rangle}).
\end{eqnarray}
These states are clearly linearly dependent and so cannot be
unambiguously discriminated at the level of one copy.  Given only
a single copy, we must tolerate a nonzero error probability in any
attempt to distinguish between these states.  If they have equal a
priori probabilities of 1/3, then the minimum error probability is
also equal to 1/3\cite{Helstrom}.  The optimum such measurement
has recently been carried out in the laboratory, where the trine
set was implemented as a set of nonorthogonal optical polarisation
states\cite{Clarke1}.  Applications of the trine set and optimal
measurements to quantum key distribution are discussed
in\cite{Phoenix}.

The trine set may be regarded as a special case of a more general
set of states having the same 3-fold rotational symmetry, but also
having a component in a third direction, which exists in a larger,
3 dimensional Hilbert space ${\cal H}_{3}\supset{\cal H}_{2}$. Let
this third dimension be spanned by the vector $|z{\rangle}$
orthogonal to both $|x{\rangle}$ and $|y{\rangle}$. This
generalised trine set may be written as
\begin{equation}
|T_{j}({\lambda}){\rangle}={\lambda}|z{\rangle}+\sqrt{1-{\lambda}^{2}}|t_{j}{\rangle},
\end{equation}
for some real parameter ${\lambda}{\in}[0,1]$ known as the {\em
lift} parameter.  When ${\lambda}=0$, the
$|T_{j}({\lambda}){\rangle}$ are just the coplanar trine states.
If, however, ${\lambda}>0$, then the states are lifted out of the
plane and are linearly independent for ${\lambda}{\neq}0,1$.
These are known as {\em lifted trine states}\cite{Shor}.  As the
lift parameter is increased, the states become increasingly
distinct until ${\lambda}=1/\sqrt{3}$, at which point they are
orthogonal. Increasing ${\lambda}$ further serves to draw the
three states closer to $|z{\rangle}$ axis until ${\lambda}=1$, at
which point $|T_{j}({\lambda}){\rangle}=|z{\rangle}$.

In this section, we show that the set of $C-$fold copies of the
trine set, which we refer to as a multi-trine set, may be
represented as a lifted trine set. We will use this, together with
the fact that the maximum discrimination probability for
lifted-trine states can be derived exactly, to determine the
maximum discrimination probability for multiple copies of the
trine states for equal a priori probabilities.

To show that the states $|t_{j}{\rangle}^{{\otimes}C}$ are lifted
trine states, we will make use of the fact that the states
$|{\tau}_{j}({\lambda}){\rangle}=|T_{j}({\lambda}){\rangle}{\otimes}|t_{j}{\rangle}$,
for ${\lambda}{\in}[0,1)$, are also lifted trine states, with a
different, nonzero lift parameter. To see this, let us define the
three orthogonal states:

\begin{eqnarray}
|X{\rangle}&=&\sqrt{\frac{2}{1+{\lambda}^{2}}}\left({\lambda}|z{\rangle}{\otimes}|x{\rangle}-\frac{\sqrt{1-{\lambda}^{2}}}{2}(|x{\rangle}{\otimes}|y{\rangle}+|y{\rangle}{\otimes}|x{\rangle})\right),
\\
*|Y{\rangle}&=&\sqrt{\frac{2}{1+{\lambda}^{2}}}\left({\lambda}|z{\rangle}{\otimes}|y{\rangle}-\frac{\sqrt{1-{\lambda}^{2}}}{2}(|x{\rangle}{\otimes}|x{\rangle}-|y{\rangle}{\otimes}|y{\rangle})\right),
\\
*|Z{\rangle}&=&\frac{1}{\sqrt{2}}(|x{\rangle}{\otimes}|x{\rangle}+|y{\rangle}{\otimes}|y{\rangle}).
\end{eqnarray}
Then the $|{\tau}_{j}({\lambda}){\rangle}$ may be written as
\begin{eqnarray}
|{\tau}_{1}({\lambda}){\rangle}&=&L|Z{\rangle}+\sqrt{1-L^{2}}|Y{\rangle},
\\
*|{\tau}_{2}({\lambda}){\rangle}&=&L|Z{\rangle}+\frac{\sqrt{1-L^{2}}}{2}(-|Y{\rangle}+\sqrt{3}|X{\rangle}),
\\
*|{\tau}_{3}({\lambda}){\rangle}&=&L|Z{\rangle}-\frac{\sqrt{1-L^{2}}}{2}(|Y{\rangle}+\sqrt{3}|X{\rangle}),
\end{eqnarray}
where the parameter $L$ is
\begin{equation}
L=\sqrt{\frac{1-{\lambda}^{2}}{2}}.
\end{equation}
Comparison of the $|{\tau}_{j}({\lambda}){\rangle}$ with the
$|T_{j}({\lambda}){\rangle}$ shows that they are indeed lifted
trine states, with lift parameter $L$, given by Eq. (3.11).  Also,
for all ${\lambda}{\in}[0,1)$, $L{\neq}0,1$ and the
$|{\tau}_{j}({\lambda}){\rangle}$ are linearly independent.

We can now use simple induction to show that the states
$|t_{j}{\rangle}^{{\otimes}C}$ are lifted trine states.  In the
above argument, if we let
$|T_{j}({\lambda}){\rangle}=|t_{j}{\rangle}$, i.e., take
${\lambda}=0$, then we find that
$|{\tau}_{j}({\lambda}){\rangle}=|t_{j}{\rangle}^{{\otimes}2}$ and
that these are lifted trine states with lift parameter
$1/\sqrt{2}$.  For the inductive step, we can say that if
$|t_{j}{\rangle}^{{\otimes}C-1}$ is a set of lifted trine states
with lift parameter $L_{C-1}$, then so is the set
$|t_{j}{\rangle}^{{\otimes}C}$, with some lift parameter $L_{C}$.
It follows from Eq. (3.11) that these lift parameters for
successive values of $C$ obey the recurrence relation
\begin{equation}
L_{C}=\sqrt{\frac{1-L_{C-1}^{2}}{2}},
\end{equation}
with the boundary condition $L_{1}=0$.  The solution is
\begin{equation}
L_{C}=\left[\frac{1}{3}\left(1-\left(\frac{-1}{2}\right)^{C-1}\right)\right]^{1/2}.
\end{equation}
So, the states $|t_{j}{\rangle}^{{\otimes}C}$ are lifted trine
states with lift parameter given by Eq. (3.13).

\subsection{Discrimination  between lifted trine states}

To determine the maximum discrimination probability for the lifted
trine set with equal a priori probabilities, we make use of the
theorem in\cite{Mesymmetric} which gives the maximum
discrimination probability for equally probable, linearly
independent symmetrical states.

A set of $N$ linearly independent symmetric states can be
expressed as
\begin{equation}
|{\psi}_{j}{\rangle}=\sum_{k=0}^{N-1}c_{k}e^{\frac{2{\pi}ijk}{N}}|u_{k}{\rangle},
\end{equation}
where $\sum_{k=0}^{N-1}|c_{k}|^{2}=1, c_{k}{\neq}0$ and
${\langle}u_{k'}|u_{k}{\rangle}={\delta}_{k'k}$.  The maximum
discrimination probability is
\begin{equation}
P_{max}=N{\times}\min_{k}|c_{k}|^{2}.
\end{equation}
For the lifted trine states, we define the following orthogonal
states:
\begin{eqnarray}
|u_{0}{\rangle}&=&|z{\rangle}, \\*
|u_{1}{\rangle}&=&\frac{e^{\frac{5{\pi}i}{6}}}{\sqrt{2}}(|x{\rangle}+i|y{\rangle}),
\\
*|u_{2}{\rangle}&=&\frac{e^{\frac{-5{\pi}i}{6}}}{\sqrt{2}}(|x{\rangle}-i|y{\rangle}).
\end{eqnarray}
In terms of these states, one can easily verify that the lifted
trine states have the form
\begin{eqnarray}
|T_{1}({\lambda}){\rangle}&=&{\lambda}|u_{0}{\rangle}+\sqrt{\frac{1-{\lambda}^{2}}{2}}\left(e^{\frac{2{\pi}i}{3}}|u_{1}{\rangle}+e^{\frac{4{\pi}i}{3}}|u_{2}{\rangle}\right),
\\
*|T_{2}({\lambda}){\rangle}&=&{\lambda}|u_{0}{\rangle}+\sqrt{\frac{1-{\lambda}^{2}}{2}}\left(e^{\frac{4{\pi}i}{3}}|u_{1}{\rangle}+e^{\frac{8{\pi}i}{3}}|u_{2}{\rangle}\right),
\\
*|T_{3}({\lambda}){\rangle}&=&{\lambda}|u_{0}{\rangle}+\sqrt{\frac{1-{\lambda}^{2}}{2}}\left(|u_{1}{\rangle}+|u_{2}{\rangle}\right).
\end{eqnarray}
One can verify that these expressions are of the form (3.14) if we
take the coefficients $c_{k}$ to be
\begin{eqnarray}
c_{0}&=&{\lambda}, \\ *
c_{1}=c_{2}&=&\sqrt{\frac{1-{\lambda}^{2}}{2}}.
\end{eqnarray}
Making use of these expressions and employing (3.15), we find that
the maximum discrimination probability for the lifted trine states
is
\begin{equation}
P_{max}=3{\times}\min\left({\lambda}^{2},\frac{1-{\lambda}^{2}}{2}\right).
\end{equation}
The behaviour of $P_{max}$ as a function of ${\lambda}$ is
illustrated in figure (1). For $0{\leq}{\lambda}{\leq}1/\sqrt{3}$,
$P_{max}=2{\lambda}^{2}$, which increases monotonically to 1 until
the orthogonality point ${\lambda}=1/\sqrt{3}$.  For
$1/\sqrt{3}{\leq}{\lambda}{\leq}1$,
$P_{max}=(3/2)(1-{\lambda}^{2})$, which decreases monotonically,
reaching zero when ${\lambda}=1$, at which point all three states
are identical.

We have shown how to calculate the maximum discrimination
probability for lifted trine states.  We will now see how these
results can be used to obtain the maximum discrimination
probability for multiple copies of the trine states.

\begin{figure}
\begin{center}
\scalebox{1}[1]{\includegraphics{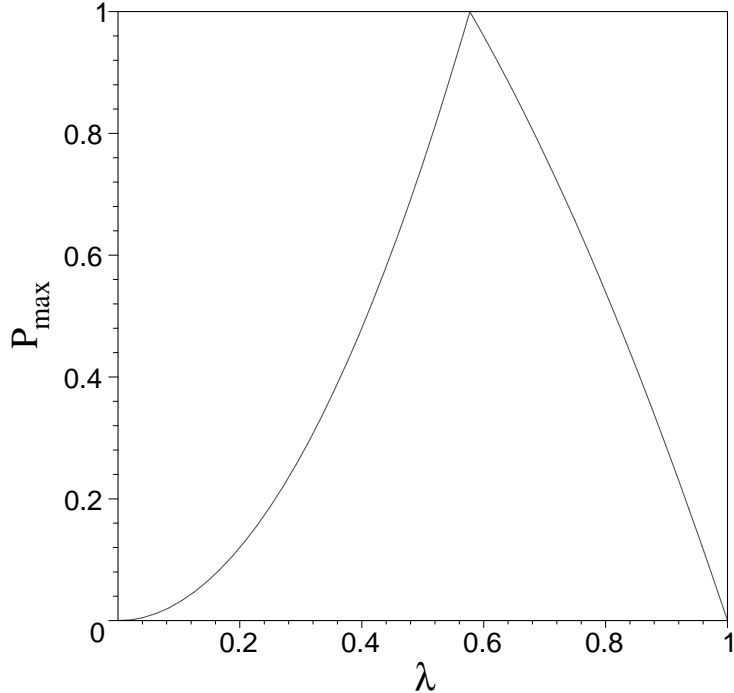}}
\end{center}

\caption{Maximum probability $P_{max}$ of unambiguous
discrimination between lifted trine states as a function of the
lift parameter ${\lambda}$.  For ${\lambda}=0,1$, the states are
linearly dependent and so unambiguous discrimination is
impossible.  However, at ${\lambda}=1/\sqrt{3}{\approx}0.557$, the
states are orthogonal and can be discriminated with unit
probability.}

\end{figure}
\subsection{Discrimination between multi-trine states}

We are now in a position to calculate the maximum discrimination
probability for the states $|t_{j}{\rangle}^{{\otimes}C}$. It
follows from (2.8) that the necessary and sufficient condition for
unambiguous discrimination is that $C{\geq}2$.   Making use of
(3.13) and (3.24), we see that the maximum discrimination
probability is
\begin{eqnarray}
P_{max}(|t_{j}{\rangle}^{{\otimes}C})&=&3{\times}\min\left(L_{C}^{2},\frac{1-L_{C}^{2}}{2}\right)
\nonumber \\ *&=&3{\times}\min\left(L_{C}^{2},L_{C+1}^{2}\right)
\nonumber
\\
*&=&\min\left(1-\left(\frac{-1}{2}\right)^{C-1},1-\left(\frac{-1}{2}\right)^{C}\right).
\end{eqnarray}
It is quite straightforward to show that the smaller of these two
terms is determined solely by whether $C$ is even or odd and we
find
\begin{equation}
P_{max}(|t_{j}{\rangle}^{{\otimes}C})=\left\{
\begin{array}{cc}
1-2^{-C} & : {\mathrm{even}}\;C \\ 1-2^{-(C-1)} & :
{\mathrm{odd}}\;C.
\end{array}
\right.
\end{equation}
Some interesting observations can be made about this result.
Firstly, the minimum probability of inconclusive results, given by
$1-P_{max}(|t_{j}{\rangle}^{{\otimes}C})$, decreases exponentially
with $C$, with even and odd cases considered separately.  However,
Eq. (3.26) has the peculiar, unexpected property that
$P_{max}(|t_{j}{\rangle}^{{\otimes}C})=P_{max}(|t_{j}{\rangle}^{{\otimes}C+1})$
for even $C$.  That is, adding another copy to an even number of
copies does not increase the maximum discrimination probability.
This behaviour provides an interesting exception to the trend
observed in state estimation/discrimination that the more copies
we have of the state, the better we can determine it\cite{Review}.

One further curious feature of the maximum discrimination
probability in Eq. (3.26) is that it can be attained by carrying
out collective discrimination measurements only on pairs of copies
of the state. Suppose that $C$ is even; if it is not, we can, in
view of the above property of
$P_{max}(|t_{j}{\rangle}^{{\otimes}C})$, simply discard one of the
copies. We divide the set of copies into $C/2$ pairs and carry out
an optimal discrimination measurement on each pair. The
probability of success for one pair is
$P_{max}(|t_{j}{\rangle}^{{\otimes}2})$. The success probability
for all $C$ copies by this method is simply the probability that
not all of the $C/2$ pairwise measurements give inconclusive
results, which is simply
$1-[1-P_{max}(|t_{j}{\rangle}^{{\otimes}2})]^{C/2}$. Into this we
insert $P_{max}(|t_{j}{\rangle}^{{\otimes}2})=3/4$, which is the
special case of (3.26) for $C=2$ and obtain the general maximum
discrimination probability in Eq. (3.26).  The ability to do
optimum discrimination for this ensemble with only pairwise
discrimination measurements is clearly convenient from a practical
perspective.

\section{Discussion}
It is impossible to discriminate unambiguously between a set of
linearly dependent states.   If, however, we have access to more
than one copy belonging to such a set, then the compound states
may be linearly independent and thus amenable to unambiguous
discrimination.  This is the possibility we explored in this
paper.

It is natural to search for any general limitations on the extent
to which this is achievable.  The most natural parameters to
consider are $D$, the dimension of the Hilbert space of a single
copy, $C$, the number of copies and $N$, the number of states.  We
derived one necessary and one sufficient condition, respectively
(2.1) and (2.3), for $N$ states to be amenable to unambiguous
discrimination for fixed $C$ and $D$. These conditions were shown
to be identical for $D=2$ and combining them, which gives (2.8),
solves the problem completely for qubits.

We then worked out in detail the specific example of unambiguous
discrimination between $C{\geq}2$ copies of the trine states. We
showed how such multi-trine states can be interpreted as lifted
trine states, for which the maximum unambiguous discrimination
probability can be calculated exactly.  We also found that if $C$
is even, then adding a further copy, strangely, fails to increase
the maximum discrimination probability.  Also, we described how
the optimum measurement for arbitrary $C{\geq}2$ can be carried
out by performing discrimination measurements only on pairs of
copies.

We conclude with an observation regarding the related subject of
probabilistic cloning.  It was established by Duan and
Guo\cite{Duanguo} that a set of quantum states can be
probabilistically copied exactly if and only if they are linearly
independent.  This result is rigorously correct for
$1{\rightarrow}M$ cloning.  If, however, $1<C<M$ copies of the
state are initially available, then sometimes $C{\rightarrow}M$
cloning will be possible for linearly dependent sets.  A
sufficient condition is the linear independence of the states
$|{\psi}_{j}{\rangle}^{{\otimes}C}$.  When this is so,
probabilistic exact cloning may be accomplished, for example, by
carrying out an unambiguous discrimination measurement to
determine the state then manufacturing $M$ copies of the state.

\section*{Appendix: Proof of Lemma}
\renewcommand{\theequation}{A.\arabic{equation}}
\setcounter{equation}{0}

{\noindent}{\bf Proof:} We prove this by contradiction. If the set
$\{|{\phi}_{k}{\rangle}{\otimes}|{\chi}_{k}{\rangle}\}\cup(|{\phi}{\rangle}{\otimes}|{\chi}{\rangle})$
is linearly dependent, then there exist coefficients $b$ and
$b_{k}$ such that
\begin{equation}
b|{\phi}{\rangle}{\otimes}|{\chi}{\rangle}+\sum_{k}b_{k}|{\phi}_{k}{\rangle}{\otimes}|{\chi}_{k}{\rangle}=0,
\end{equation}
where not all of the coefficients in $\{b,b_{k}\}$ are zero. In
fact, we can show that at least two of the $b_{k}$ are nonzero. If
only one of the $b_{k}$ were nonzero, then the corresponding
$|{\chi}_{k}{\rangle}$ would be equal to either $|{\chi}{\rangle}$
(up to a phase) or the zero vector, depending on whether or not
$b=0$.  The latter possibility contradicts the premises of the
lemma (normalisation). The former does also, since it would imply
that for the nonzero $b_{k}$, $|{\chi}_{k}{\rangle}$ is not
distinct from $|{\chi}{\rangle}$.

The set $\{|{\phi}_{k}{\rangle}\}$ is linearly independent, so
there exists a set of reciprocal states $\{|{\tilde
\phi}_{k}{\rangle}\}{\in}{\cal H}'$ such that ${\langle}{\tilde
\phi}_{k'}|{\phi}_{k}{\rangle}={\langle}{\tilde
\phi}_{k}|{\phi}_{k}{\rangle}{\delta}_{kk'}$ and ${\langle}{\tilde
\phi}_{k}|{\phi}_{k}{\rangle}{\neq}0\;{\forall}\;k$. Acting on Eq.
(A.1) throughout with ${\langle}{\tilde \phi}_{k}|{\otimes}1$
gives
\begin{equation}
b{\langle}{\tilde
\phi}_{k}|{\phi}{\rangle}|{\chi}{\rangle}+b_{k}{\langle}{\tilde
\phi}_{k}|{\phi}_{k}{\rangle}|{\chi}_{k}{\rangle}=0\;{\forall}\;k.
\end{equation}
The fact that at least two of the $b_{k}$ are nonzero implies that
the corresponding $|{\chi}_{k}{\rangle}$ will be indistinct,
contradicting the premise.  This completes the proof.

\section*{Acknowledgements}
The author would like to thank Vivien M. Kendon for a conversation
in which the issue addressed in this paper arose. This work was
funded by the UK Engineering and Physical Sciences Research
Council.


\begin{thebibliography}{99}
\bibitem{CCapacity}
A. S. Holevo, {\it IEEE Trans. Inf. Theory} {\bf IT-44} 269
(1998); B. Schumacher and M. Westmoreland, {\it Phys. Rev. A} {\bf
56} 131 (1997).
\bibitem{NWE}
C. H. Bennett, D. P. DiVincenzo, C. A. Fuchs, T. Mor, E. Rains, P.
W. Shor, J. A. Smolin and W. K. Wootters, {\em Phys. Rev. A} {\bf
59} 1070 (1999).
\bibitem{Review}
A. Chefles, {\it Contemp. Phys.} {\bf 41} 401 (2000).
\bibitem{Linear}
A. Chefles, {\em Phys. Lett. A} {\bf 239} 339 (1998).
\bibitem{Shor}
P. W. Shor, `On the number of elements needed in a POVM attaining
the accessible information', LANL eprint quant-ph/0009077.
\bibitem{Bhatia}
R. Bhatia, {\it Matrix Analysis}, (Springer-Verlag, Berlin, 1991).
\bibitem{Broida}
See, for example, J. G. Broida and S. G. Williamson, {\it A
Comprehensive Introduction to Linear Algebra}, (Addison-Wesley,
1989).
\bibitem{Helstrom}
C. W. Helstrom, {\em Quantum Detection and Estimation Theory},
(Academic Press, New York, 1976).
\bibitem{Clarke1}
R. B. M. Clarke, V. M. Kendon, A. Chefles, S. M. Barnett, E. Riis
and M. Sasaki, {\em Phys. Rev. A} {\bf 64} 012303 (2001).
\bibitem{Phoenix}
S. J. D. Phoenix, S. M. Barnett and A. Chefles, {\em J. Mod. Opt.}
{\bf 47} 507 (2000).
\bibitem{Mesymmetric}
A. Chefles and S. M. Barnett, {\em Phys. Lett. A} {\bf 250} 223
(1998).
\bibitem{Duanguo}
L. M. Duan and G. C. Guo, {\em Phys. Rev. Lett.} {\bf 80} 4999
(1998).

\end{thebibliography}
\end{document}